\documentclass{iau}
\usepackage{graphicx}
\usepackage{multirow}
\usepackage{textcomp}
\usepackage{epsfig}
\usepackage{amsmath}

\def\aj{\ref@jnl{AJ}}                   
             
\def\apj{{\textit{ApJ}}}

\def\sgrf{SGR\,0418$+$5729}

\newcommand{\chandra}{{\it Chandra}}

\newcommand{\xmm}{{\em XMM--Newton}}

\newcommand{\suzaku}{{\em Suzaku}}
\newcommand{\rxte}{{\em RXTE}}
\newcommand{\swift}{{\em Swift}}

\newcommand{\bc}{\begin{center}}
\newcommand{\ec}{\end{center}}
\def\ltsima{$\; \buildrel < \over \sim \;$}
\def\lsim{\lower.5ex\hbox{\ltsima}}
\def\loe{\lower.5ex\hbox{\ltsima}}
\def\gtsima{$\; \buildrel > \over \sim \;$}
\def\gsim{\lower.5ex\hbox{\gtsima}}
\def\goe{\lower.5ex\hbox{\gtsima}}

\def\ltsima{$\; \buildrel < \over \sim \;$}
\def\lsim{\lower.5ex\hbox{\ltsima}}
\def\loe{\lower.5ex\hbox{\ltsima}}
\def\gtsima{$\; \buildrel > \over \sim \;$}
\def\gsim{\lower.5ex\hbox{\gtsima}}
\def\goe{\lower.5ex\hbox{\gtsima}}

\def\ergs {erg\,s$^{-1}$}
\def\ergscm2 {erg\,s$^{-1}$cm$^{-2}$}
\def\ss {s\,s$^{-1}$}
\def\cm2 {cm$^{-2}$}

\def\ergs {${\rm erg\, s}^{-1}$}

\def\src {Swift\,J1822.3--1606}

\title[IAUS291.~~A new low-B magnetar: Swift\,J1822.3--1606]{A new low-B magnetar: Swift\,J1822.3--1606} 

\author[A. Camero-Arranz  et al.]  
{A. Camero-Arranz$^1$, N. Rea$^1$, G. L. Israel$^2$, P. Esposito$^3$, J. A. Pons$^4$, 
R. P. Mignani$^{5,6}$, R. Turolla$^{7,5}$, S. Zane$^5$, 
M. Burgay$^3$, A. Possenti$^3$, \\ 
S. Campana$^8$,  
T. Enoto$^9$, 
N. Gehrels$^{10}$, E. G{\"o}{\u g}{\"u}{\c s}$^{11}$, D. G\"otz$^{12}$, \\C. Kouveliotou$^{13}$,
K. Makishima$^{14,15}$, 
S. Mereghetti$^{16}$ and S. R. Oates$^5$}

\affiliation{$^1$Institut de Ci\`encies de l'Espai (CSIC-IEEC), Campus UAB, Facultat de Ci\`encies,\\ 
Torre C5-parell, E-08193 Barcelona, Spain \\ email: {\tt camero@ice.cat} \\[\affilskip] 
$^2$INAF, Osservatorio Astronomico di Roma, via Frascati 22, 00040, Monteporzio Catone, Italy\\[\affilskip] 
$^3$INAF, Osservatorio Astronomico di Cagliari, Poggio dei Pini, I-09012 Capoterra, Italy\\[\affilskip] 
$^4$Departament de Fisica Aplicada, Univ. d'Alacant, Ap. Correus 99, 03080 Alacant, Spain\\[\affilskip] 
$^5$Mullard Space Science Laboratory, Univ. College London, Dorking, Surrey RH5 6NT, UK\\[\affilskip] 
$^6$Institute of Astronomy, University of Zielona G\'ora, Lubuska 2, 65-265, Zielona G\'ora, Poland\\[\affilskip] 
$^7$Universit\`a di Padova, Dipart. di Fisica e Astronomia, via F.~Marzolo 8, I-35131 Padova, Italy\\[\affilskip] 
$^8$INAF -- Osservatorio Astronomico di Brera, via E.~Bianchi 46, I-23807 Merate, Italy\\[\affilskip] 
$^9$Kavli Institute for Particle Astrophysics \& Cosmology (KIPAC), SLAC/Stanford University, PO Box 20450, MS 29, Stanford, CA 94309, USA\\[\affilskip] 
$^{10}$NASA Goddard Space Flight Center, Greenbelt, MD 20771, USA\\[\affilskip] 
$^{11}$Sabanc\i\ University, Orhanl\i-Tuzla, 34956 \.Istanbul, Turkey\\[\affilskip] 
$^{12}$AIM (CEA/DSM-CNRS-Universit\'e Paris Diderot), Irfu/Service d'Astrophysique, Saclay, F-91191 Gif-sur-Yvette, France\\[\affilskip] 
$^{13}$NASA Marshall Space Flight Center, Huntsville, AL 35812, USA\\[\affilskip] 
$^{14}$High Energy Astrophysics Laboratory, Institute of Physical and Chemical Research (RIKEN), Wako, Saitama 351-0198, Japan\\[\affilskip] 
$^{15}$Department of Physics, Univ. of Tokyo, 7-3-1 Hongo, Bunkyo-ku, Tokyo 113-0033, Japan\\[\affilskip] 
$^{16}$INAF -- IASF Milano, via E. Bassini 15, I-20133 Milano, Italy\\[\affilskip] }

\pubyear{2012}
\volume{291}  
\jname{\mbox{Neutron Stars and Pulsars: Challenges and Opportunities after 80 years}}
\editors{J. van Leeuwen, ed.} 

\begin{document}

\maketitle


\begin{abstract}

We report on the long term X-ray monitoring with \swift, \rxte,
\suzaku, \chandra, and \xmm\ of the outburst of the newly discovered magnetar
\src\,(SGR\,1822-1606), from the first observations soon after the
detection of the short X-ray bursts which led to its discovery (July 2011),
through the first stages of its outburst decay (April  2012).  Our X-ray timing
analysis finds the source rotating with a period of
$P=8.43772016(2)$\,s and a period derivative
$\dot{P}=8.3(2)\times10^{-14}$~\ss , which entails an inferred dipolar
surface magnetic field of $B\simeq2.7\times10^{13}$~G at the equator. This
measurement makes \src\, the second lowest magnetic field magnetar (after \sgrf; Rea et al. 2010). Following
the flux and spectral evolution from the beginning of the outburst, 
we find that the flux decreased by about an order of
magnitude, with a subtle softening of the spectrum, both typical of
the outburst decay of magnetars. By modeling the secular thermal
evolution of \src, we find that the observed timing properties of the
source, as well as its quiescent X-ray luminosity, can be reproduced if it
was born with a poloidal and crustal toroidal fields of
$B_{p}\sim1.5\times10^{14}$\, G and $B_{tor}\sim7\times10^{14}$\, G,
respectively, and if its current age is $\sim$550\,kyr (Rea et al. 2012).

\keywords{ stars: magnetic fields --- stars: neutron --- X-rays: \src}

\end{abstract}

\firstsection 
\section{X-ray spectral modeling}\label{spectral}

In this study, we used all available data obtained from different space-based satellites, covering a time-span
from July 2011 until end of April  2012.  Spectra were extracted for all the \rxte/PCA, \swift/XRT, \suzaku/XIS03,
and \xmm/pn\, data, using standard software provided by the different
team missions, and modeled using XSPEC version 12.7.0. Best fits were
found using a blackbody plus power-law (BB+PL; $\chi^2_\nu/$dof
=1.05/2522) and a 2 blackbodies (2BBs; $\chi^2_\nu/$dof =1.06/2522)
model, all corrected for the photoelectric absorption. Figure 1
(left) shows how the flux decreased by about an order of magnitude, typical of the outburst decay of magnetars.

\begin{figure}[!t]
\includegraphics[trim=0 12 0 30, clip, height=6cm, width=0.5\textwidth]{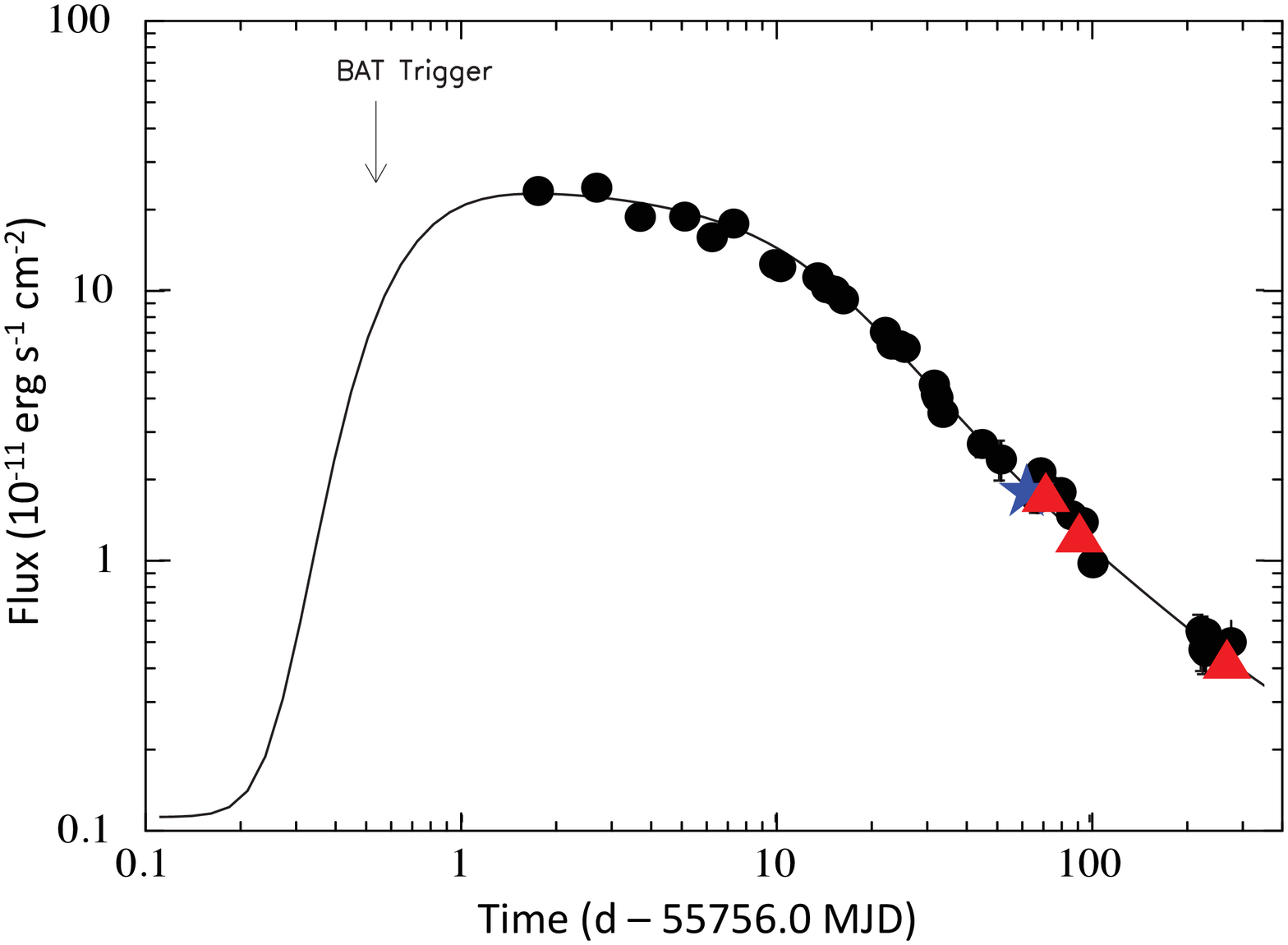}
\includegraphics[trim=10 25 230 95, clip,  height=6cm, width=0.5\textwidth]{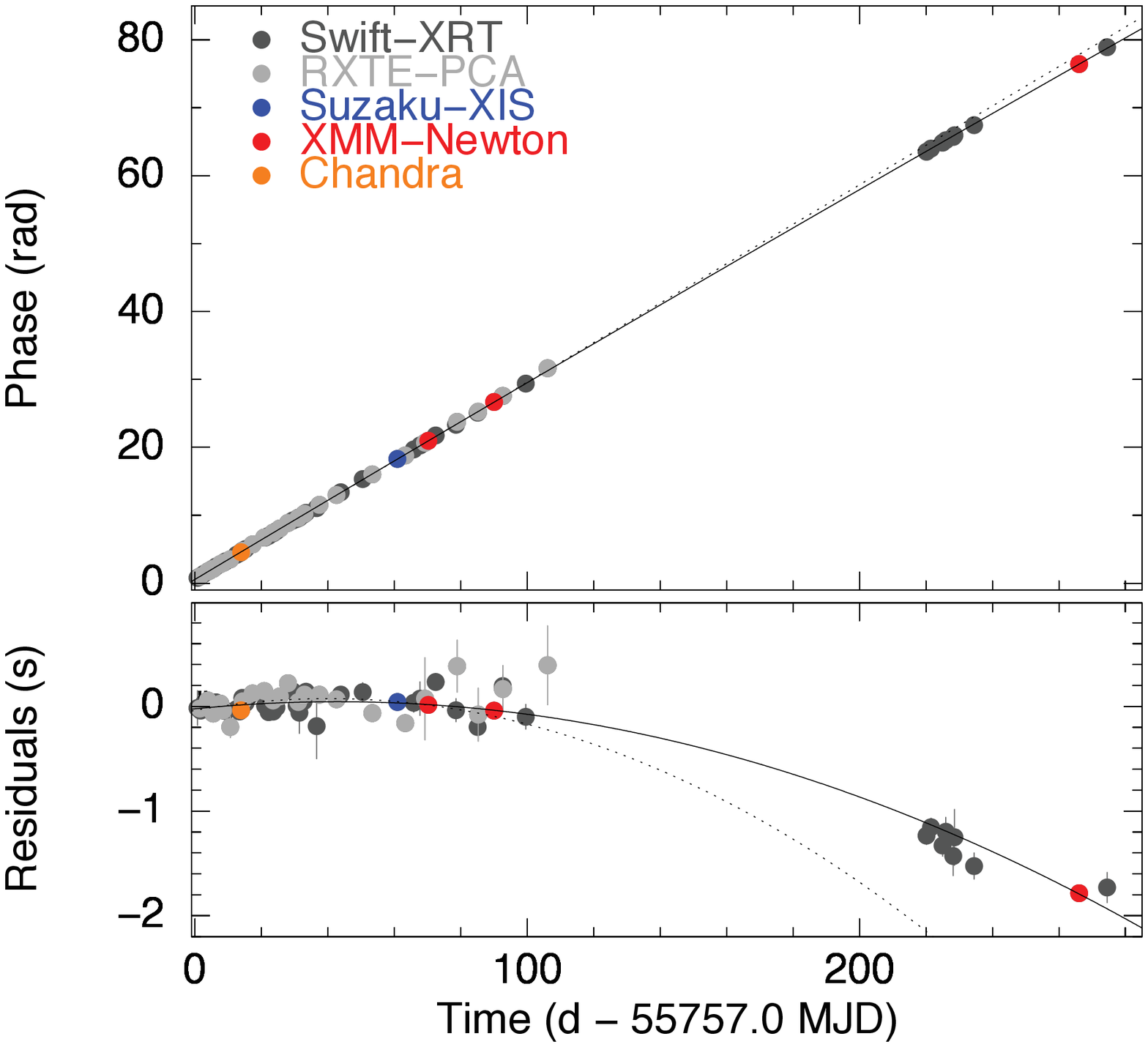}
\caption{{\em Left:} Outburst model from \cite[Pons \& Rea (2012)]{pons12} superimposed to the 1-10\,keV flux decay of \src\,. Black circles denote \swift/XRT data, red triangles correspond to \textit{XMM-Newton}  and blue stars to \suzaku/XIS03\, data. 
{\em Right:} Pulse phase evolution as a function of time, together with the time residuals (lower panel) after having corrected for the linear component (correction to the $P$ value). The solid lines in the two panels mark the inferred  $P$--$\dot{P}$ coherent solution based on the whole dataset, while the dotted lines represent the $P$--$\dot{P}$ coherent solution based on the data collected during the first 90 days only.}
\label{fig:phases}
\end{figure}

The aggressive monitoring campaign we present here allowed us not only to study
in detail the flux decay of \src, but also to give an estimate of its typical
timescale.  We have compared the observed outburst decay with the more physical theoretical model presented in Pons \&  Rea (2012).  In addition, we have performed numerical
simulations with a 2D code designed to model the magneto-thermal evolution of neutron stars. In Figure 1 (left), super-imposed,  we show our best representative model that reproduce the observed properties of the decay of \src\, outburst.  This model corresponds to an injection of $4\times10^{25}$~erg~cm$^{-3}$ in the outer crust, in the narrow layer with density between $6\times10^{8}$ and $6\times 10^{10}$\,g~cm$^{-3}$, and in an angular region of 35 degrees (0.6 rad) around the pole. The total injected energy was then $1.3\times10^{42}$~erg. 

\section{X-ray timing analysis}\label{timing}

For the X-ray timing analysis we used all available data after  barycentering all the events.  
We started by obtaining an accurate period measurement by folding the data from the first two XRT pointings which were separated by less than 1\,day, and studying the phase evolution within these observations by means of a phase-fitting technique (see \cite[Dall'Osso et~al.\ 2003]{dallosso03} for details).  The resulting best-fit period (reduced $\chi^2=1.1$ for 2 dof) is $P=8.43966(2)$\,s (all errors are given at 1$\sigma$ c.l.) at the epoch MJD 55757.0. The above period accuracy of 20\,$\mu$s is enough to phase-connect coherently the later \swift, \rxte, \chandra, \suzaku, and \xmm\, pointings (see Figure\,\ref{fig:phases}).  

We modeled the phase evolution with a linear plus quadratic term. The corresponding coherent solution (valid until November 2011) is  $P=8.43772007(9)$\,s and period derivative $\dot{P} = 1.1(2)\times 10^{-13}$\,s s$^{-1}$ ($\chi^2=132$ for 57 dof; at epoch MJD 55757.0). The above solution accuracy allows us to unambiguously extrapolate the phase evolution until the beginning of the next \swift\ visibility window which started in February 2012. The final resulting phase-coherent solution, once the latest 2012 observations are included, returns a best-fit period of $P=8.43772016(2)$\,s and period derivative of $\dot{P} = 8.3(2)\times 10^{-14}$\,s~s$^{-1}$  at MJD 55757.0 ($\chi^2=145$ for 67 dof). The above best-fit values imply a surface dipolar magnetic field of $B\simeq2.7\times10^{13}$\,G (at the equator), a characteristic age of $\tau_{\rm c}=P/2\dot{P}\simeq1.6$\,Myr, and a spin-down power L$_{\rm rot}=4\pi I \dot{P}/P^3\simeq1.7\times 10^{30}$ \ergs (assuming a neutron star radius of 10\,km and a mass of 1.4$M_{\odot}$). The final solution has a relatively high r.m.s. ($\sim$ 120\,ms) resulting in a best-fit reduced $\chi_{\nu}^2=2.1$.  The 3$\sigma$ upper limit of the second derivative of the period was $\ddot{P}<5.8\times10^{-21}$s~s$^{-2}$ (but see also Livingstone et al. 2011 and Scholz et al. 2012). 

\section{Conclusions}

We have reported on the outburst evolution of the new magnetar \src, which,
despite its relatively low magnetic field ($B=2.7\times10^{13}$\,G),
is in line with the outbursts observed for other magnetars with
higher dipolar magnetic fields.

We  found that the current properties of the source can be reproduced if it has now an age of $\sim550$\,kyr, and it was born with a toroidal crustal field of $7\times10^{14}$\, G, which has by now decayed by less than an order of magnitude.

The position of \src\ in the $P$--$\dot P$ diagram is close to that of the ``low'' field magnetar
\sgrf\, (\cite[Rea et al.\ 2010]{rea10}).  As argued in more detail in Rea
et al. (2012), we note  that the discovery of a second magnetar-like source with a magnetic field in
the radio-pulsar range strengthens the idea that magnetar-like behavior may be much more widespread than what believed in the past, and that it is related to the intensity and topology of the internal and surface toroidal components, rather than only to the surface dipolar field (\cite[Rea et al.\ 2010]{rea10}, \cite[Perna \& Pons\ 2011]{perna11}, \cite[Turolla et al.\ 2011]{turolla11}).

\end{document}